# Carbon Nanotube-based Super Nanotube: Tailorable Thermal Conductivity at Three-dimensional

*Haifei Zhan, John M. Bell and Yuantong Gu\**

*School of Chemistry, Physics and Mechanical Engineering, Queensland University of Technology, 2 George St, Brisbane QLD 4001, Australia*

**\*Corresponding Author:** Professor Yuantong Gu

**Mailing Address:** School of Chemistry, Physics and Mechanical Engineering,

Queensland University of Technology,

GPO Box 2434, Brisbane, QLD 4001, Australia

**Telephones:** +61-7-31381009          **Fax:** +61-7-31381469

**E-mail:** yuantong.gu@qut.edu.au

**Abstract**

The advancements of nanomaterials/nanostructures have enabled the possibility of fabricating multifunctional materials that hold great promises in engineering applications. The carbon nanotube (CNT)-based nanostructure is one representative building block for such multifunctional materials. Based on a series of *in silico* studies, we report the tailorability of the thermal conductivity of a three-dimensional CNT-based nanostructure, i.e., the single wall CNT (SWNT)-based super nanotube (ST). It is shown that the thermal conductivity of STs varies with different connecting carbon rings, and the ST with longer constituent SWNTs and larger diameter yield to a smaller thermal conductivity. Further results reveal that the inverse of the ST's thermal conductivity exhibits a good linear relationship with the inverse of its length. Particularly, it is found that the thermal conductivity exhibits an approximately proportional relationship with the inverse of the temperature, but appears insensitive to the axial strain due to its Poisson's ratio. These results, in the one hand, provide a fundamental understanding of the thermal transport properties of the super carbon nanotubes conductivities of ST, and in the other hand shed lights on their future design/fabrication and engineering applications.

1. Introduction

The advancement of nanotechnology has enabled the continuing miniaturization of devices and also the possibility of fabricating multifunctional materials that show great promises in a wide range of engineering applications, such as adaptive airfoils, robotic skins.[1] The nanoelectromechanical systems (NEMS) that could response to an extremely small shift of mass or force due to the integrated sensors and electronic circuits,[2] presenting an infancy stage of such multifunctional material. As the building blocks for those multifunctional materials, the low dimensional carbon materials, such as carbon nanotube (CNT) and graphene, have received intensive interests from both scientific and engineering communities. For instance, the atomic-thin graphene shows a great promise for the applications of portable and wearable energy conversion and storage devices (e.g., fuel cells, supercapacitors).[3] Particularly, the versatile flexibility of carbon in forming different hybridization states allows the fabrication of various carbonaceous nanomaterials/nanostructures from zero-dimensional (0D) to 3D, realizing different desired functionalities.[4] For example, the thermal transport properties of 2D carbon networks can be effectively altered through the change of its structure.[5] The recently synthesized 3D CNT pillared graphene structure[6] shows broad prospects for applications as thermal sinker and electronic devices.[7]

Among various carbon-based nanostructures, the CNT-based nanostructures have received the most intensive interests, such as the 3D CNT-graphene hybrid nanostructures[8] and 2D super graphene[9] and 3D super carbon nanotube.[10] Particularly, the super carbon nanotube (ST) is constructed by replacing the C-C bond with a single wall CNT (SWNT). Researchers have reported that the ST exhibits metallic and semiconducting behaviour,[11] high flexibility,[12] ultra-high sensitivity as mass sensor ($\sim 10^{-24}$ g).[13] Majority of current studies have focused on the mechanical properties of the ST, e.g., the mechanical behaviours under tension,[14] compression and bending.[15] To facilitate various promising applications of the ST, like fuel cells, battery electrodes or nanoelectronic devices, a comprehensive understanding of its thermal transport properties is a prerequisite, which however is lacking currently.

Previous studies have shown that the strength, stiffness and toughness of ST can be optimized simultaneously, benefiting from the high flexibility while constructing its structure.[16] Herein, we employ large-scale molecular dynamics (MD) simulations to assess the thermal conductivities of the ST with changing geometrical structures, sizes, temperature, and strain status. The obtained results are expected in the one hand to establish a basic understanding of the thermal conductivities of ST, and in the other hand to guide its design and applications.

2. **Computational Methods**

The thermal conductivity ($\kappa$) of the ST is calculated based on the Muller-Plathe's method,[17] known as the reverse nonequilibrium molecular dynamics simulations (rNEMD). Before the calculation, the configuration of the ST is firstly optimized by the conjugate gradient minimization method, and then equilibrated using Nose-Hoover thermostat[18-19] under ambient conditions for 500 ps (temperature = 300 K and pressure = 1 atm). Afterwards, a constant heat flux ($J$) is introduced to the system through a velocity exchange scheme under the microcanonical ensemble, i.e., NVE ensemble with constant atom number, volume, and energy. $\kappa$ is then calculated when the system arrives the steady state regime according to $\kappa = -J/(\partial T/\partial x)$. Here $\partial T / \partial x$ is the temperature gradient along the heat flux direction ($x$), and $J$ is the heat flux computed from

$$J = \frac{1}{2tA} \sum_N \frac{m}{2}(v_{hot}^2 - v_{cold}^2) \tag{1}$$

Here, $t$ is total simulation time, $N$ is the total number of exchanges, $m$ is the atomic mass, $v_{hot}$ and $v_{cold}$ are the exchange velocities of the hot and cold atoms, respectively. $A$ is the cross-sectional area. The factor 2 in the denominator is used to account for the periodicity of the system. To ensure a reliable thermal conductivity, $\kappa$ is calculated at a time interval of 250 ps for a total simulation time of 4 ns and averaged over the simulation duration from 3 to 4 ns. During the whole simulation, a small time step of 0.5 fs is chosen, and the C-C atomic interactions were modelled by the widely used adaptive intermolecular reactive empirical bond order (AIREBO) potential,[20] which has been shown to well represent the binding energy and elastic properties of carbon materials. Periodic boundary conditions were applied along the length direction of the ST. The whole calculation is performed under the software package LAMMPS.[21]

## 3. Results and Discussions

### 3.1 Structural Influence

At the beginning, we acquire the thermal conductivity of the ST with different structures. As illustrated in Fig. 1, different architectures can be obtained by altering the constituent SWNTs (both radius and length), the chirality of the ST, and also the atomic configurations at the junctions. Fig. 1b illustrates the top view of three different junction types (denoted as α, β and γ, respectively), in which three constituent zigzag (8,0) CNTs are connected by six heptagons, six adjacent heptagons, and six pairs of pentagon and octagon, respectively. For discussion convenience, the ST that is constructed from (n, m) SWNTs is denoted as ST(N, M)@(n, m)-α (α representing the junction type). Thus, the radius of the ST can be calculated from $R_{ST} = l_{cnt}\sqrt{3(N^2 + MN + M^2)}/(2\pi)$, where $l_{cnt}$ is the distance between the adjacent junctions (see Fig. 1a).[14] That is, the outer and inner radii of the ST equal to $R_{ST} + d_{cnt}/2$ and $R_{ST} - d_{cnt}/2$ ($d_{cnt}$ is the diameter of the constituent SWNTs), respectively. The initially discussed three ST models are comprised by (8,0) SWNTs, which share a similar size with a same $l_{cnt}$ ~19 Å. Due to the highly porous nature of the ST, we approximate its cross-section as a continuous annulus, i.e., the cross-sectional area is approximated as $2\pi R_{ST} d_{cnt}$ while calculating the heat flux from Eq. 1. To note that such approximation should not influence our discussion as we will focus on the relative thermal conductivity of various STs.

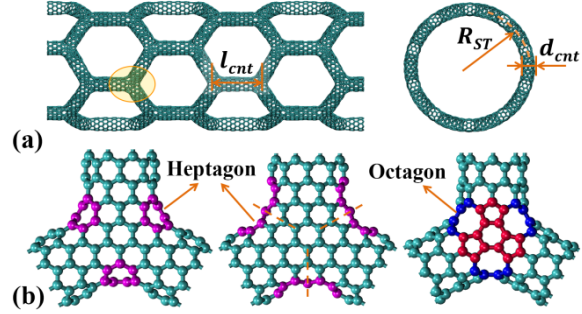

**Fig. 1** (a) The atomic configuration of a ST; (b) Three different junctions constructed by (8,0) SWNT, denoted as (8,0)-α, (8,0)-β, and (8,0)-γ from left to right, respectively. The connecting pentagons, heptagons and octagons are highlighted by red, magenta and blue, respectively.

In general, a relatively low thermal conductivity is estimated for different STs, e.g., $\kappa$ for the ST(5,0)@(8,0)-α is around 5.4 ± 0.1 W/mK. Such low thermal conductivity is majorly stemmed from the continuum approximation of the highly porous structure of the ST, which will greatly increase the cross-sectional area (Fig. 1a). Refer to Eq. 1, the much larger cross-sectional area of the ST comparing with a SWNT will greatly reduce the calculated thermal conductivity (see Supplementary Information for a more detailed discussion). Apparently, from the left panel of Fig. 2a, the junction type plays an important role in the thermal transport properties of the ST. For all three junction types (i.e., ST(5,0)@(8,0), ST's diameter of ~6 nm and length of ~75 nm), the effective $\kappa$ of the ST with β junction is ~36% larger than its counterpart with α junction. The origins of such difference can be evidenced by looking at their vibrational density of states (VDOS, computed from the Fourier transformation of the autocorrelation function of the atomic velocities[22]). As seen in Fig. 2b, the VDOS spectrum varies with the junction type, which therefore endows the ST with various thermal conductivities. Generally, the ST exhibits a similar but shifted pattern of VDOS as comparing to a SWNT, i.e., low-frequency (out-of-plane) and high-frequency (in-plane) phonon modes around 20 and 60 THz, respectively. However, unlike a SWNT where the amplitude of the high-frequency component is much larger than that of the low-frequency, the ST shows similar amplitude for the low- and high-frequency components. Such observation indicates that the out-of-plane phonons also play an important role in the thermal transport properties of the ST due to its 3D structure.

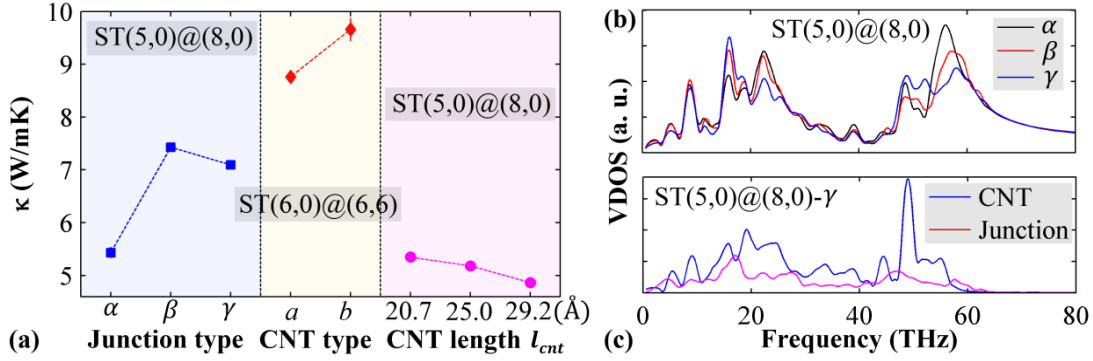

**Fig. 1** (a) Thermal conductivity of different STs, the errorbar represents the standard deviation of the thermal conductivity during the simulation time of 3 to 4 ns; (b) Comparisons of the VDOS between different junction types; (c) Comparisons of the VDOS at the CNT and junction regions.

We then look at the thermal conductivity of ST with a similar geometry size but different constituent CNTs (i.e., ST(6,0)@(6,6), see the Supplementary Information for the atomic configurations). The ST(6,0)@(6,6) has a $l_{cnt}$ ~15 Å, with an outer radius about 29 Å, similar as the radius of the ST(5,0)@(8,0). As illustrated in the middle panel of Fig. 2a, the effective $\kappa$ of the two STs constructed from armchair (6,6) SWNTs are much larger than that of the ST made from (8,0) SWNTs. For instance, the effective $\kappa$ of the ST(6,0)@(6,6)-*b* is more than 80% larger than the $\kappa$ of the ST(5,0)@(8,0)-α. This observation suggests the tunability of the thermal conductivity of the ST through altering the charity of the constituent CNTs. Two origins are assumed to be responsible for the big difference of the thermal conductivity between the ST(5,0)@(8,8) and ST(6,0)@(6,6). Firstly, different constituent CNTs require different connecting carbon rings, which will induce different VDOS as compared in Fig. 2b, and therefore results in different thermal transport properties. As shown in Fig. 2c, there are evident mismatches of the VDOS between the CNT and junction regions, and such VDOS mismatches are determined by the connecting carbon rings at the junctions. Secondly, different constituent CNT means different geometry sizes of the ST, including the cross-sectional thickness and radius. This geometry difference exerts significant impacts on the thermal transport properties of the ST as discussed in below section. In the right panel of Fig. 2a, the effective $\kappa$ for STs with the same constituent CNT, similar ST's length, but different CNT's length ($l_{cnt}$) are compared. Although the ST with longer constituent CNTs has fewer junctions, i.e., less phonon scatterings, a lower thermal conductivity is observed. This phenomenon can also be explained by considering the impact from the ST's diameter as discussed in the below section, i.e., the ST's diameter increases with the increasing constituent CNT's length.

*3.2 Size and Temperature Dependence*

Besides the evident geometrical impacts on the thermal conductivity of the ST as discussed above, $\kappa$ is also supposed to be sensitive to the sample size (length and radius). According to the common heuristic argument that when the periodic length of the simulation cell $L_{CNT}$ is smaller than the phonon mean-free path $\Lambda$ ($\Lambda$ ~700-750 nm for SWNT at room temperature[23]), the inverse of thermal conductivity is proportional to the frequency of scattering events, with contributions from the sample ends and from intrinsic scattering,[24-25] i.e., $1/\kappa \propto 1/\Lambda + 1/L_{CNT}$. Consistent with such assumption, the inverse of $\kappa$ exhibits a good linear relationship with the inverse of the ST's length as illustrated in Fig. 3a. By extrapolating the linear trend to $L$~∞, the limit of thermal conductivity for a macroscopic ST $\kappa_\infty$ can be obtained. From Figure 3a, the limit of thermal conductivity for the ST(5,0)@(8,0)-γ is estimated as $\kappa_\infty = 5.9$ W/mK, i.e., $\kappa$ will saturate to $\kappa_\infty$ at 300 K while the sample size is much larger than the phonon mean-free path.

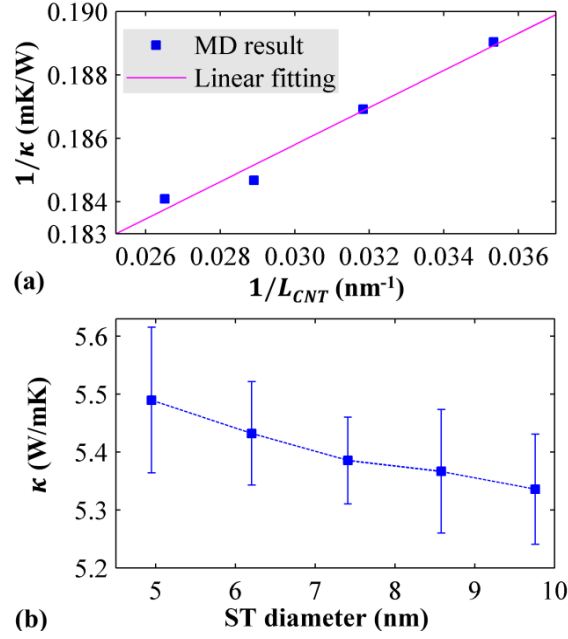

**Fig. 2** Results obtained from different ST(N,0)@(8,0)-γ: (a) the inverse of thermal conductivity as a function of the inverse of the ST's length for ST(5,0)@(8,0)-γ; (b) the thermal conductivity as a function of the ST's diameter (outer diameter).

Meanwhile, the dependence of the effective $\kappa$ on the ST's diameter has also been examined. As shown in Fig 3b, the effective $\kappa$ decreases gradually and continuously with the increasing ST's diameter, which is in line with the results reported for SWNT.[26] Theoretically, the heat in the ST is majorly transferred by lattice vibration owing to the strong covalent $sp^2$ bonds, i.e., the lattice thermal conductivity at a given temperature is expressed as[27]

$$\kappa(T) = \sum_{j} C_j(T) v_j^2 \tau_j(T) \qquad (2)$$

Here $C_j$, $v_j$ and $\tau_j$ are the specific heat, phonon group velocity and phonon relaxation time of the phonon mode $j$ (including two transverse acoustic modes and one longitudinal acoustic mode), respectively. According to the Matthiessen rule,[26] the phonon relaxation time is mainly determined by boundary scattering and the three-phonon umklapp scattering processes, i.e., $1/\tau = 1/\tau_b + 1/\tau_u$. Here $\tau_b$ and $\tau_u$ represent the relaxation-time parameters for boundary scattering and three-phonon umklapp scattering process, respectively. Similar to the SWNT,[26] the increase of ST's diameter is assumed to decrease the average group velocity and also augment the probability of the umklapp process, which will thus degrade the thermal conductivity in the ST. This observation also explains the lower $\kappa$ observed for the ST with longer constituent SWNT as presented in the right panel of Fig. 2a. That is the enhancement of the thermal transport properties from the less junction phonon scattering (originated from the longer constituent SWNT) is much smaller than the degrading effect resulted from the increased ST's diameter. Therefore, the ST with longer constituent SWNTs exhibits smaller $\kappa$. We should note that the observed geometry size dependent thermal conductivity also explains the big difference of $\kappa$ estimated between the similar "super" ST in Fig. 2a, i.e., ST(6,0)@(6,6) and ST(5,0)@(8,0).

Additionally, we also assessed the thermal conductivity of the ST under different temperatures. From Fig. 4, the effective $\kappa$ is approximately proportional to the inverse of the temperature, which is consistent with the results reported for SWNTs.[27-28] It is expected that as the temperature increases, the higher-energy phonons are thermally populated which will enhance the role of umklapp scattering in determining $\kappa$ (i.e., $\tau_u$ will dominate the overall phonon relaxation time $\tau$). According to Eq. 2, while the umklapp process dominates the phonon process in a ST, the corresponding relaxation rates increases rapidly, which will thus result in a continuously decreasing $\kappa$ as observed in Fig. 4.

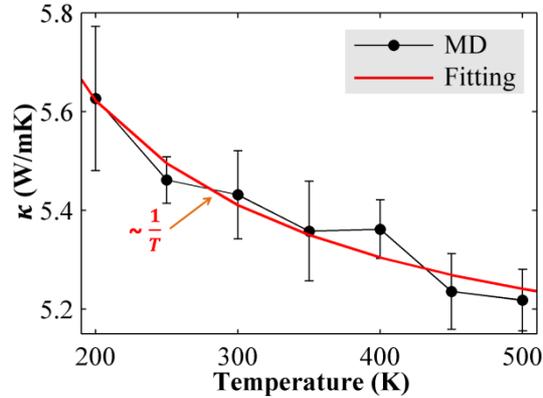

**Fig. 3** Effective thermal conductivity of the ST(5,0)@(8,0)-γ as a function of temperature.

*3.3 Strain Influence*

Before concluding, we examined the strain impacts on the thermal transport properties of the ST. A constant strain rate of 0.0002 ps$^{-1}$ was adopted to elongate/compress the ST to achieve a strain from -2% (compression) to 5% (tension) after relaxation. To note that the ST is still in the elastic deformation regime within this strain range (see

Supplementary Information). Same simulations are performed to estimate the thermal conductivity of the strained ST.

Fig. 5 shows the relative $\kappa$ of the ST(5,0)@(8,0)-γ and a (8,0) SWNT as a function of axial strain. As is seen, the relative $\kappa$ of ST fluctuates around one, indicating insignificant impacts from the axial strain on its thermal conductivity. Whereas, for a (8,0) SWNT, the increasing strain from negative (compression) to positive (tension) induces a significant reduction to the thermal conductivity. Such observations are consistent with the VDOS results as shown in Fig. 5b and 5c. For the SWNT, the axial stain softens the high-frequency phonon modes (i.e. G-band phonon modes), which decreases the specific heat and thus leads to a continuous reduction to $\kappa$ (from compressive strain to tensile strain). However, for the ST, the low-frequency and high-frequency phonon modes are almost unchanged at different strains. Such facts yield to the phenomenon that the strain exerts marginal influence to the thermal conductivity. More specifically, for the ST under axial strain, its constituent SWNTs will be either compressed or stretched depending on their orientations. Take the tensile stain as an example, the constituent SWNT along the loading direction will under tensile deformation while the SWNT along the lateral direction will under compressive deformation due to the Poisson's ratio (see inset of Fig. 5a). In other words, the enhancement and degrading effect from the compressed and stretched SWNTs will compensate to each other and thus makes the thermal conductivity of the ST insensitive to the axial strain.

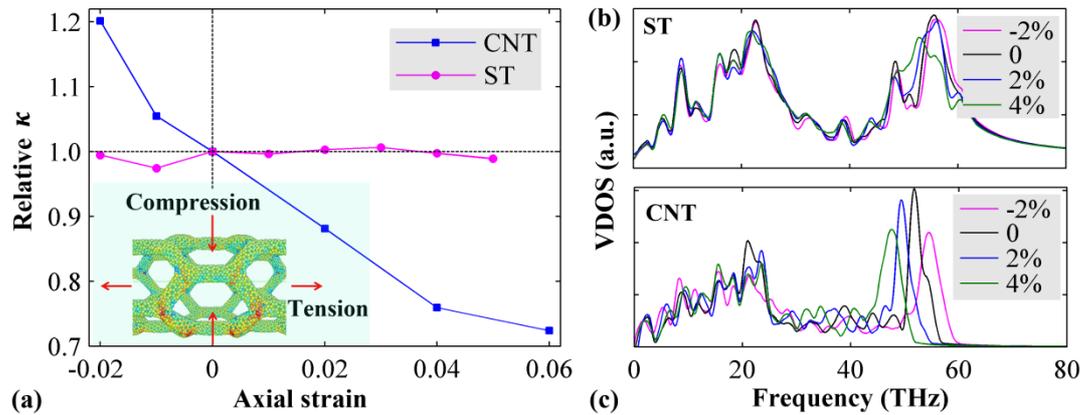

**Fig. 4** (a) Relative thermal conductivity of the ST(5,0)@(8,0)-γ and SWNT as a function of axial strain. Comparisons of the VDOS at different axial strains for: (b) ST and (c) SWNT.

**4. Conclusions**

Based on a series of *in silico* studies, we investigated the tailorability of the CNT-based super nanotube's (ST's) thermal conductivity through altering its geometrical structure, size, temperature, and strain. It is found that the STs with different connecting carbon rings possess different VDOS spectrums, which yield to different thermal conductivity. Further studies show that the ST with longer constituent SWNTs and larger diameter exhibits smaller thermal conductivity, which is considered as a result from the decreased average group velocity and augmented probability of the umklapp process. Consistent with previous studies, the inverse of thermal conductivity is proportional to the frequency of scattering events, i.e., the inverse of ST's $\kappa$ exhibits a good linear relationship with the inverse of its length. Further, the effective $\kappa$ shows an approximately proportional relationship with the inverse of the temperature, i.e., $\kappa \propto 1/T$. Additional tests reveal that the thermal conductivity of the ST is insensitive to the axial strain due to its Poisson's ratio, i.e., due to compensations of the enhancement and degrading effect from the simultaneously compressed and stretched SWNTs under axial strain. This study provides a fundamental understanding of the thermal transport properties of the super carbon nanotubes, which should shed lights on their future design/fabrication and engineering applications.

**Acknowledgement**

Supports from the ARC Discovery Project (DP130102120) and the High Performance Computer resources provided by the Queensland University of Technology are gratefully acknowledged. Dr Zhan is grateful to Dr Zhou from Hefei University of Technology (China) for providing the advices to construct the models.

**Supplementary Information**

Supplementary information includes the atomic configurations of ST's unit cells made from (6,6) SWNTs, and the stress-strain curves of the ST under axial strain.

## Table of Contents

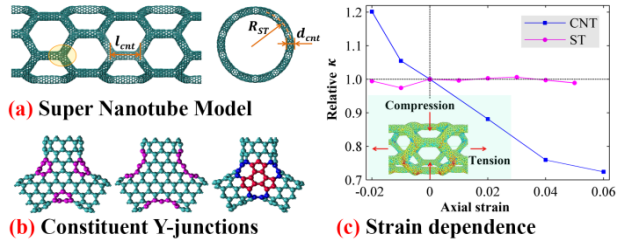

A first numerical study on the tailorability of the thermal conductivity of three-dimensional CNT-based nanotube.

# Supplementary Information

**Carbon Nanotube-based Super Nanotube: Tailorable Thermal Conductivity at Three-dimensional**

*Haifei Zhan, John M. Bell and Yuantong Gu\**

*School of Chemistry, Physics and Mechanical Engineering, Queensland University of Technology, 2 George St, Brisbane QLD 4001, Australia*

**S1.** Schematic view of the ST's cross-section.

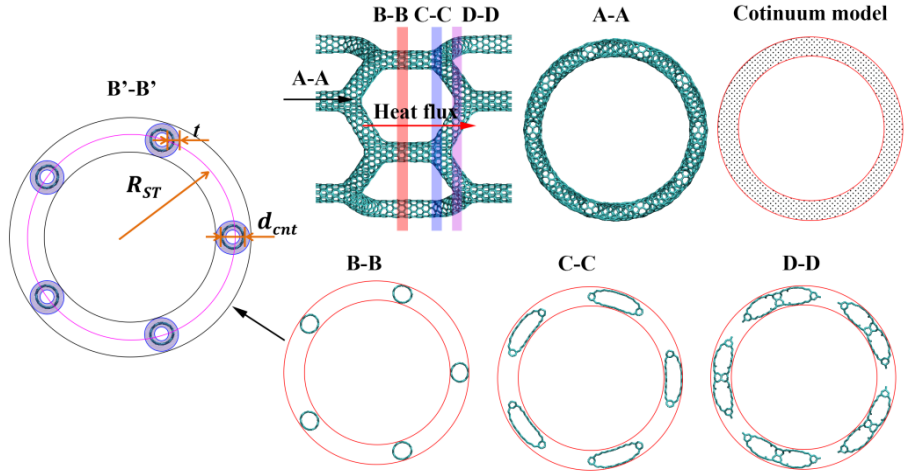

Above figure clearly shows that the cross-section of the ST changes along the heat flux direction (the B-B, C-C and D-D view of the cross-section). Due to the highly porous nature of the structure, it is hard to get an exact cross-sectional area. Therefore, current work adopted a continuum model to estimate the cross-sectional area, which will greatly increase the cross-sectional. Although such approximation will lead to an extremely small thermal conductivity value, it will not influence the discussions as we focus on the relative thermal conductivity of various ST.

A simple example is given to show how the estimation of the cross-sectional area will impact the calculated thermal conductivity. As illustrated in the B'-B' view, suppose that the constituent CNT has a thickness $t = 0.34\ nm$, then the cross-sectional area of the continuum and discrete (B'-B') model can be estimated as $A1 = 2\pi R_{ST}(d_{cnt} + t)$, and $A2 = 5\pi d_{cnt}t$, respectively. Thus, we have $A_r = \frac{A1}{A2} = \frac{2R_{ST}(d_{cnt}+t)}{5d_{cnt}t}$. Recall the ST(5,0)@(8,0)-γ model with $2R_{ST} \sim 6\ nm$, $d_{cnt} \sim 0.63\ nm$, $A_r$ is estimated around 5.4. In other words, the absolute thermal conductivity would be 5.4 larger if we adopt the discrete (B'-B') model to estimate the cross-sectional area. It is evident from such approximation that larger ST's radius will lead to bigger difference for the estimated thermal conductivity between the continuum model and the discrete model (B'-B').

**S2.** Atomic configurations of the junction units ST(6,0)@(6,6)-a and ST(6,0)@(6,6)-b.

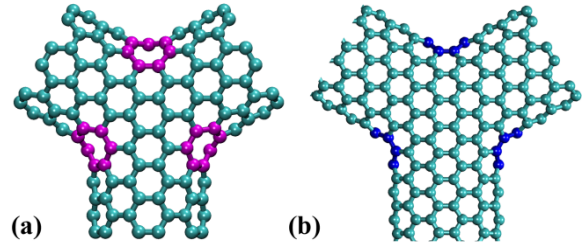

(a)    (b)

**S3.** Stress-strain curves of the ST under tension (T) and compression (C) using different potential cut-off distances.

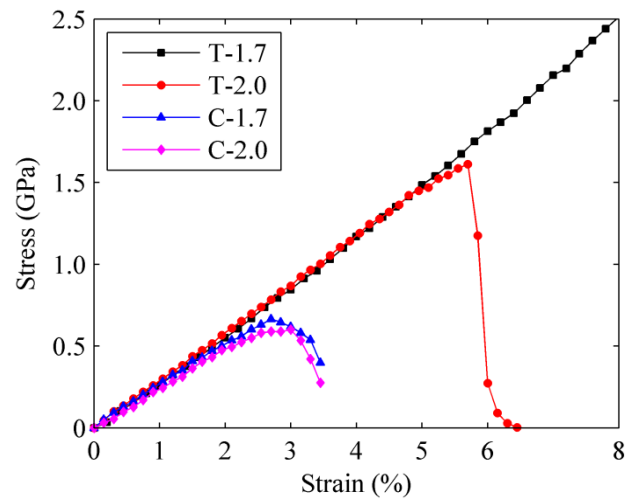

To examine the strain influence, the ST was elongated along length direction under a constant strain rate of 0.0002 ps$^{-1}$. To notice that the original AIREBO potential would result in spuriously high tensile force when the C-C bonds are stretched beyond 1.7 Å. Researchers have suggested that a 2.0 Å cutoff distance can be chosen to overcome such abnormal phenomenon.[1] As compared in above figure, the chosen of different cutoff distance will not alter the elastic deformation of the ST, and the emphasis of this paper is the impacts from the elastic strain. Thus, the original cutoff distance of 1.7 Å is chosen to achieve a pre-strained structure for the consistency with the thermal conductivity calculation.